\documentclass[%
 reprint,
 superscriptaddress,
 nofootinbib,
 twocolumn, 
 amsmath,amssymb,
 aps,
prb,
]{revtex4-2}

\usepackage[colorlinks = true,
            linkcolor = blue,
            urlcolor  = red,
            citecolor = red,
            anchorcolor = blue]{hyperref}
\usepackage[T1]{fontenc}
\usepackage[utf8]{inputenc}
\usepackage{graphicx}% Include figure files
\usepackage{dcolumn}% Align table columns on decimal point
\usepackage{color}
\usepackage{amsmath}
\usepackage{appendix}
\usepackage{MnSymbol}
\usepackage{orcidlink}
\usepackage{todonotes}
\usepackage{glossaries}
\usepackage{ulem}
\newacronym{CDW}{CDW}{charge-density-wave}
\newacronym{RIXS}{RIXS}{resonant inelastic x-ray scattering}
\newacronym{DMRG}{DMRG}{density matrix renormalization group}
\newacronym{BOW}{BOW}{bond-order wave}
\newacronym{DQMC}{DQMC}{determinant quantum Monte Carlo}
\newacronym{QMC}{QMC}{quantum Monte Carlo}
\newacronym{SSH}{SSH}{Su-Schrieffer-Heeger}
\newacronym{oSSH}{oSSH}{optical Su-Schrieffer-Heeger}
\newacronym{2D}{2D}{two-dimensional}
\newacronym{1D}{1D}{one-dimensional}
\newacronym{FS}{FS}{Fermi surface}
\newacronym{eph}{$e$-ph}{electron-phonon}
\newacronym{HMC}{HMC}{hybrid Monte Carlo}
\newacronym{KVB}{KVB}{Kekul{\'e} Valence Bond}
\newacronym{KVBS}{KVBS}{Kekul{\'e} Valence Bond Solid}
\newacronym{VBS}{VBS}{Valence Bond Solid}
\newacronym{SM}{SM}{semi-metal}
\newacronym{QCP}{QCP}{quantum critical point}
\newacronym{FSS}{FSS}{finite-size scaling}
\newacronym{AFM}{AFM}{antiferromagnetic}

\newacronym{SC}{SC}{superconducting}

\begin{document}

\preprint{}
\title{Antiferromagnetism and Kekul{\'e} valence bond order in the honeycomb optical Su-Schrieffer-Heeger-Hubbard model}

\author{Sohan~{Malkaruge Costa}\orcidlink{0000-0002-9829-9017}}
\affiliation{Department of Physics and Astronomy, The University of Tennessee, Knoxville, Tennessee 37996, USA}
\affiliation{Institute for Advanced Materials and Manufacturing, The University of Tennessee, Knoxville, Tennessee 37996, USA\looseness=-1} 

\author{Benjamin~Cohen-Stead\orcidlink{0000-0002-7915-6280}}
\affiliation{Department of Physics and Astronomy, The University of Tennessee, Knoxville, Tennessee 37996, USA}
\affiliation{Institute for Advanced Materials and Manufacturing, The University of Tennessee, Knoxville, Tennessee 37996, USA\looseness=-1} 

\author{Steven~Johnston\orcidlink{0000-0002-2343-0113}}
\affiliation{Department of Physics and Astronomy, The University of Tennessee, Knoxville, Tennessee 37996, USA}
\affiliation{Institute for Advanced Materials and Manufacturing, The University of Tennessee, Knoxville, Tennessee 37996, USA\looseness=-1}

\date{\today}% It is always \today, today,

\begin{abstract}
The precise role of \gls*{eph} coupling in graphene and related materials on a honeycomb lattice is not yet fully understood, despite extensive research on these systems. Here, we perform sign-problem-free determinant quantum Monte Carlo simulations of the \gls*{oSSH}-Hubbard model on the honeycomb lattice, focusing on the parameters relevant to graphene. Performing finite-size scaling analyzes, we obtain the model's ground state phase diagram, which includes the \gls*{SM}, \gls*{KVBS}, and \gls*{AFM} phases, as well as indications of a small \gls*{KVBS}/\gls*{AFM} coexistence region.
We find that a weak to moderate Hubbard repulsion, tuned toward the \gls*{SM}–\gls*{AFM} critical value in the pure honeycomb Hubbard model, enhances \gls*{KVBS} correlations and can even stabilize the \gls*{KVBS} phase.
Estimating the effective parameters for graphene places it in the \gls*{SM} region of the phase diagram, but near the \gls*{SM}-\gls*{KVBS} phase boundary. Notably, we predict that increasing either the on-site Hubbard  repulsion or the \gls*{eph} coupling strength drives graphene toward the \gls*{KVBS} phase rather than the \gls*{AFM} phase, highlighting a  synergistic effect that can be exploited to further control the remarkable properties of graphene and related materials. 
\end{abstract}

% This command resets the glossery package so acronyms are again spelt out in the main text.
\glsresetall

\maketitle

\noindent\textbf{Introduction}.  
The synthesis of single-layer graphene~\cite{Novoselov2004Electric}, and the subsequent discovery of superconductivity in magic-angle twisted bilayer graphene~\cite{Cao2018Unconventional}, has motivated extensive investigations of \gls*{2D} honeycomb systems~\cite{Herbut2009Theory, Assaad2013Pinning, Sentef2015theory, Toldin2015Fermionic, Zhang2019Charge, Chen2019Charge, Costa2021Magnetism, Weber2021Valence, Liu2024electron, Dashwood2021probing, Malkaruge2024Kekule, Otsuka2024Kekule, Kennedy2025Extended, Andrade2025Topical}. These studies have focused on topics ranging from Dirac fermion physics~\cite{Novoselov2005Two},  fractionalized quantum Hall effects~\cite{Bolotin2009Observation}, Moir{\'e} superlattices~\cite{Cao2018Unconventional}, topological phases~\cite{Ren2016Topological}, and the response to applied mechanical effects such as strain~\cite{Naumis2023Mechanical, Sorella2018Correlation, Ribeiro2009Strained, Eom2020Direct}. A variety of models have subsequently been proposed to describe the low-energy physics of graphene-based systems. For example, a series of studies has addressed the nature and location of a \gls*{QCP} in the honeycomb Hubbard model separating its \gls*{SM} and \gls*{AFM} phases~\cite{Sorella2012Absence,Assaad2013Pinning,Toldin2015Fermionic,Kennedy2025Extended}. There have also been studies investigating the emergence of \gls*{CDW} order in the honeycomb Holstein model for \gls*{eph} interactions~\cite{Zhang2019Charge, Chen2019Charge} and its competition with \gls*{AFM} order when local Hubbard interactions are reintroduced~\cite{Costa2021Magnetism}. 

An example of an experimentally observed emergent phenomena thought to involve both \gls*{eph} interactions and electronic correlations is the formation of Kekul{\'e}-O lattice distortions in strained graphene systems~\cite{Andrade2025Topical, Eom2020Direct, Sorella2018Correlation}. Several models incorporating electronic interactions give rise to a \gls*{KVBS} ground state consistent with the Kekul{\'e}-O order~\cite{Li2017Rermion, Xu2018Kekule, Li2023Emergent}. However, this phase is also accompanied by a corresponding lattice distortion~\cite{Zhang2022Self}, implying that the bond-stretching phonons may also play a role. This scenario is supported by angle-resolved photoemission experiments that have inferred strong \gls*{eph} coupling to graphene's bond-stretching optical phonons~\cite{Leem2009highresolution, Mazzola2013kinks, Ma2019direct, Zhang2022Self}. Given that \gls*{SSH}-like \gls*{eph} interactions have been linked to the emergence of \gls*{VBS} phases in various contexts~\cite{Mazumdar2009from, Li2010paired, Weber2021Valence, Cai2021antiferromagnetism, Goetz2022valence, Malkaruge2023Comparative, Tanjaroon2023Comparative, Tanjaroon2025Antiferromagnetic,Casebolt2025Magnetic,Silva2025Two}, including in the presence of competing electronic correlations~\cite{Feng2022Phase}, it is reasonable to expect they might also play an important role in the formation of Kekul{\'e}-O lattice distortions. Recent studies have shown just this, revealing that a \gls*{KVBS} phase emerges in honeycomb \gls*{SSH} models~\cite{Malkaruge2024Kekule, Otsuka2024Kekule}. However, these studies either neglected electronic correlations entirely~\cite{Malkaruge2024Kekule} or only included the \gls*{eph} interaction at the mean-field level in the adiabatic limit~\cite{Otsuka2024Kekule}. 

Here we investigate the effects of both local Hubbard and \gls*{oSSH} interactions on a honeycomb lattice while treating both interactions on an equal footing. Specifically, we perform numerically-exact \gls*{DQMC} simulations to study the competition between emergent \gls*{KVBS} and \gls*{AFM} correlations and map out the ground-state phase diagram as a function of the Hubbard and \gls*{eph} interaction strength at half-filling. Unlike previous studies, our \gls*{DQMC} calculations simulate fully quantum mechanical phonons, while also treating the electronic correlations exactly. We find that a Hubbard $U \lesssim 3.8t$ enhances \gls*{KVBS} correlations and lowers the critical \gls*{eph} coupling needed to stabilize this phase. We further find indications of a small \gls*{KVBS}-\gls*{AFM} coexistence region, which may reflect the presence of a weakly first-order phase transition between the two phases. By considering parameter values relevant for graphene, our results suggest that both $e$-ph and Hubbard-like interactions play a role in the emergence of Kekul{\'e}-O lattice distortions in strained graphene systems.
\\

\noindent\textbf{Model}.  
The \gls*{oSSH}-Hubbard Hamiltonian on a honeycomb lattice can be conveniently partitioned as $\hat{H} = \hat{H}_e + \hat{H}_{\text{ph.}}+\hat{H}_{e-\text{ph.}}+\hat{H}_{e-e}$. The first term
\begin{equation}
    \hat{H}_e = -t\sum_{\boldsymbol{i},\nu,\sigma}\left[\hat c_{\boldsymbol{i}+\boldsymbol{r}_\nu,\mathrm{B},\sigma}^\dagger \hat c_{\boldsymbol{i},\mathrm{A},\sigma}^{\phantom\dagger}+\text{h.c.}\right]-\mu\sum_{\boldsymbol{i},\gamma,\sigma} \hat{n}_{\boldsymbol{i},\gamma,\sigma}
\end{equation}
is a noninteracting electron tight-binding Hamiltonian. Here $\hat c_{\boldsymbol{i},\gamma,\sigma}^\dagger \ (\hat c_{\boldsymbol{i},\gamma,\sigma}^{\phantom\dagger})$ creates (annihilates) a spin-$\sigma$ electron on orbital $\gamma\in ({\text{A},\text{B}})$ of unit-cell $\boldsymbol{i}$ in the lattice, with $ \hat{n}_{\boldsymbol{i},\gamma,\sigma}^{\phantom\dagger} = \hat{c}_{\boldsymbol{i},\gamma,\sigma}^{\dagger} \hat{c}_{\boldsymbol{i},\gamma,\sigma}^{\phantom\dagger}$ the corresponding number operator. The sum over $\boldsymbol{r}_\nu$ ($\nu = 0, 1, 2$) runs over the three displacement vectors connecting an $\text A$ orbital to its nearest-neighbors on sublattice $\text B$. The parameters $t$ and $\mu$ are nearest-neighbor hopping amplitude and the chemical potential, respectively. Throughout this work, we consider a half-filled particle-hole symmetric system ($\mu=0$), where the \gls*{FS} consists of a pair of Dirac points at $\boldsymbol{K}_\pm=\frac{2\pi}{a}\left(\frac 1 3,\pm\frac{1}{3\sqrt 3}\right)$ and $a$ is the lattice constant.

The second term in $\hat{H}$ introduces the noninteracting lattice (phonon) degrees of freedom
\begin{equation}
    \hat{H}_{\text{ph.}} = \sum_{\boldsymbol{i},\gamma}\left[\frac{\hat P_{\boldsymbol{i},\gamma}^2}{2M}+\frac{1}{2}M\Omega^2\hat R_{\boldsymbol{i},\gamma}^2\right]. 
\end{equation}
Here, we place two dispersionless optical phonon modes on each orbital to describe the in-plane motion of ions. Each phonon mode has frequency $\Omega = 0.1t$. The momentum and displacement operators for the pair of modes on each orbital are given by $\hat{\boldsymbol{P}}_{\boldsymbol{i},\gamma}=\left(\hat P_{\boldsymbol{i},\gamma,X},\hat P_{\boldsymbol{i},\gamma,Y}\right)$ and $\hat{\boldsymbol{R}}_{\boldsymbol{i},\gamma}=\left(\hat X_{\boldsymbol{i},\gamma},\hat Y_{\boldsymbol{i},\gamma}\right)$, respectively, with $\hat{P}_{\boldsymbol{i},\gamma}=|\hat{\boldsymbol{P}}_{\boldsymbol{i},\gamma}|$ and $\hat{R}_{\boldsymbol{i},\gamma} =|\hat{\boldsymbol{R}}_{\boldsymbol{i},\gamma}|$ and are defined such that the ion mass $M = 1$.

The third term in $\hat{H}$ describes the \gls*{eph} coupling 
\begin{equation}
    \hat{H}_{e-\text{ph.}}=\alpha\sum_{\boldsymbol{i},\nu,\sigma}\Delta \hat R_{\boldsymbol i,\nu}\left[ \hat c_{\boldsymbol{i}+\boldsymbol{r}_\nu,B,\sigma}^\dagger \hat c_{\boldsymbol{i},A,\sigma}^{\phantom\dagger}+\text{h.c.}\right],
\end{equation}
where $\Delta \hat R_{\boldsymbol i,\nu} = \left(\hat{\boldsymbol{R}}_{\boldsymbol i + \boldsymbol r_\nu,B}-\hat{\boldsymbol{R}}_{\boldsymbol i,A}\right)\cdot \boldsymbol r_\nu /|\boldsymbol r_\nu|$ is the relative displacement of phonon modes on neighboring orbitals projected into the equilibrium bond direction. The \gls*{eph} interaction arises from an \gls*{SSH}-like coupling~\cite{Barisic1970tightbinding, su1980soliton, Capone1997small}, where the lattice displacements linearly modulate the electron hopping energy, with the parameter $\alpha$ controlling the strength of the coupling. In the remainder of this paper we will report results in terms of a dimensionless \gls*{eph} coupling constant $\lambda = \alpha^2/\left(M\Omega^2t \right)$.

Finally, the last term is the local Hubbard repulsion
\begin{equation}
    \hat{H}_{e-e} = U\sum_{\boldsymbol i,\gamma} (\hat{n}_{\boldsymbol i,\gamma,\uparrow}-\tfrac{1}{2}) (\hat{n}_{\boldsymbol i,\gamma,\downarrow}-\tfrac{1}{2}),
\end{equation}
where $U>0$ controls the interaction strength.
\\

\noindent\textbf{Methods}.  
We solve the model using \gls*{DQMC} simulations with \gls*{HMC} updates to efficiently sample the phonon fields~\cite{Stefan2018Revisiting,Benjamin2022Fast,Ostmeyer2025First,Ostmeyer2025Minimal}, as implemented in \texttt{SmoQyDQMC.jl} package~\cite{Benjamin2024SmoQyDQMC.jl,Benjamin2024Codebase}. Because we are focusing on half-filling, the model has particle-hole symmetry and the \gls*{DQMC} simulations are sign-problem-free~\cite{Tanjaroon2025Antiferromagnetic}. Additional details of the \gls*{DQMC} simulations are provided in App.~\ref{sec:dqmc_details}

Based on previous work~\cite{Malkaruge2024Kekule, Otsuka2024Kekule, Kennedy2025Extended, Costa2021Magnetism}, we expect the model to potentially host \gls*{SM}, \gls*{AFM}, and \gls*{KVBS} phases. The \gls*{KVBS} phase breaks a $C_3$ lattice-rotation symmetry while the \gls*{AFM} phase breaks $\text{SU}(2)$ spin-rotation symmetry. Their respective order parameters are given by
\begin{equation}
    \hat\Psi_\text{VBS}(\boldsymbol{Q})=\frac{1}{L^2}\sum_{\boldsymbol{j},\nu}e^{-\mathrm{i}\cdot\boldsymbol{Q}\cdot\boldsymbol{j}} e^{-\mathrm{i}2\pi\nu/3}\hat{B}_{\boldsymbol{j},\nu}
\end{equation}
and
\begin{equation}
    \hat\Psi_\text{AFM}(\boldsymbol{Q})=\frac{1}{L^2}\sum_{\boldsymbol{j}}e^{-\mathrm{i}\boldsymbol{Q}\cdot\boldsymbol{j}}\left(\hat S^z_{\boldsymbol{j},\mathrm{A}}-S^z_{\boldsymbol{j},\mathrm{B}}\right),
\end{equation}
where $\hat B_{\boldsymbol{j},\nu}=\sum_\sigma \left(\hat c_{\boldsymbol{j}+r_\nu,\mathrm{B},\sigma}^\dagger\hat c^{\phantom\dagger}_{\boldsymbol{j},\mathrm{A},\sigma} + \text{h.c.}\right)$ and $\hat S_{\boldsymbol{i},\gamma}^z=\tfrac{1}{2}(\hat{n}_{\boldsymbol{j},\gamma,\uparrow}-\hat{n}_{\boldsymbol{j},\gamma,\downarrow})$ are the local bond and spin-$z$ operators, respectively, and $L$ the linear extent of the cluster in the direction of each lattice vector. The corresponding structure factors are $S_{\theta}(\boldsymbol{Q})=L^2|\hat{\Psi}_{\theta}(\boldsymbol{Q})|$, where $\theta = \text{AFM}\text{ or }\text{VBS}$ denotes the type of order. The ordering wave-vectors for \gls*{KVBS} and \gls*{AFM} order are $\boldsymbol{Q}=\boldsymbol{K}_+$ and $\boldsymbol{\Gamma}$, respectively.

For finite-size scaling, we also measure the corresponding correlation ratio $R_\theta = 1 - \tfrac{1}{6}\sum_{\delta \boldsymbol{q}}S_{\theta}(\boldsymbol{Q}+\delta \boldsymbol{q})/S_{\theta}(\boldsymbol{Q})$
associated with each structure factor, where $\delta\boldsymbol{q}$ denotes the six nearest-neighbor momentum points to the ordering wave-vector. Then, assuming Lorentz invariance, we take $\beta = Lt$ to project onto the ground-state~\cite{Herbut2009Theory,Assaad2013Pinning,Otsuka2016Universal,Toldin2015Fermionic}. The crossing points of the correlation ratio curves as a function of the coupling for sequentially adjacent lattice sizes then provide an estimate of the critical coupling and can be used to locate a \gls*{QCP}. 
\\

\begin{figure}[t]
    \centering
    \includegraphics[width=1.0\linewidth]{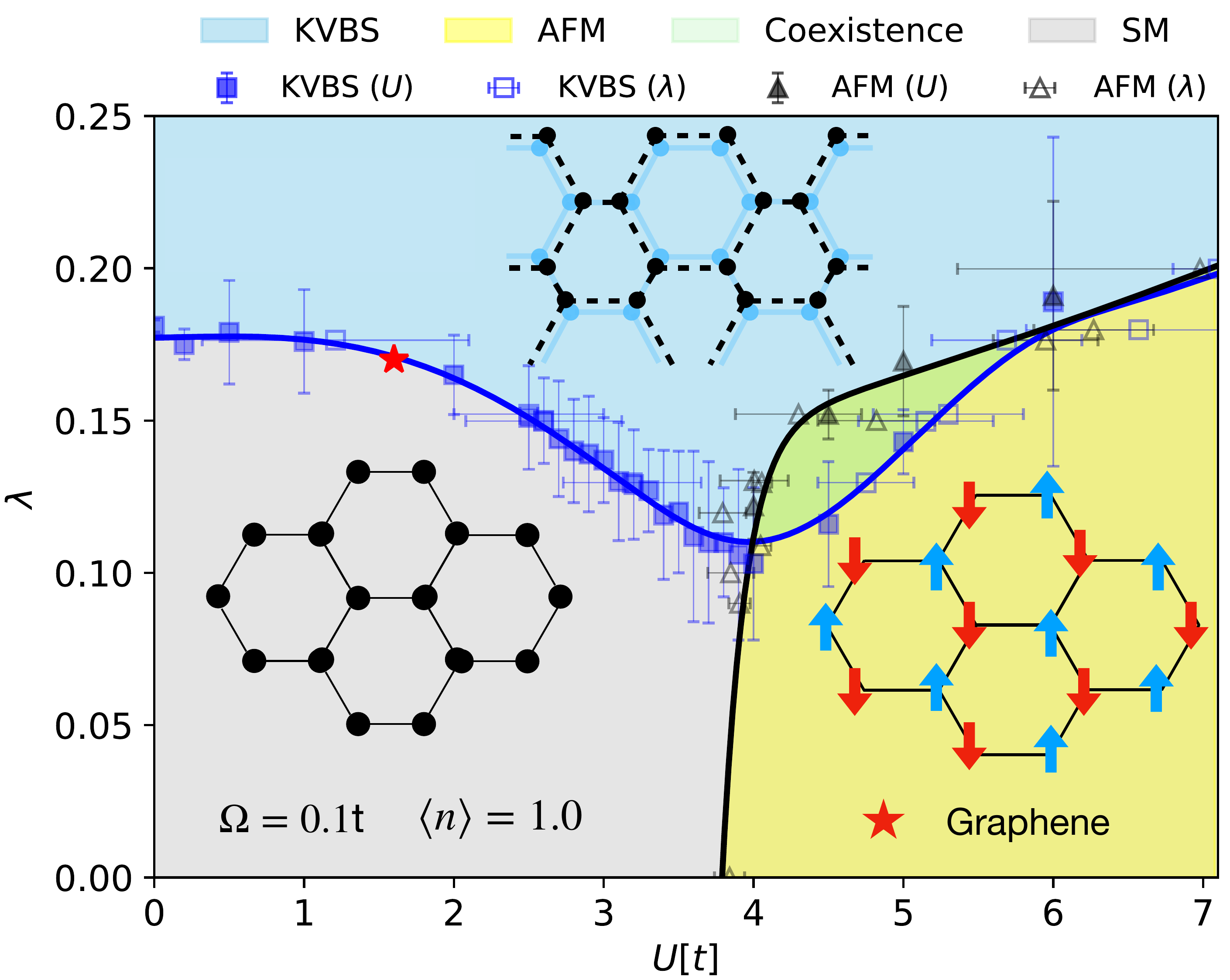}
    \vspace{-0.4cm}
    \caption{Ground state $U-\lambda$ phase diagram for the Hubbard-oSSH model on a half-filled ($\langle n \rangle = 1$) honeycomb lattice, obtained for $\Omega=0.1t$. Blue and black markers indicate a QCP for Kekul{\'e} valence bond solid (KVBS) and antiferromagnetic (AFM) phase transitions, respectively. Filled (empty) markers indicate the transition point found for fixed $U \ (\lambda)$ to determine $\lambda_\mathrm{c} \ (U_\mathrm{c})$ for the relevant phase transition. Refer to Fig.~\ref{fig:corr_ratio_vs_couplings} for examples of this analysis. The red star mark indicates the approximate location of graphene in the phase diagram. The blue and black lines are guides to the eye.} 
    \label{fig:Phase_diagram}
\end{figure}

\begin{figure}[t]
    \centering
    \includegraphics[width=1.0\linewidth]{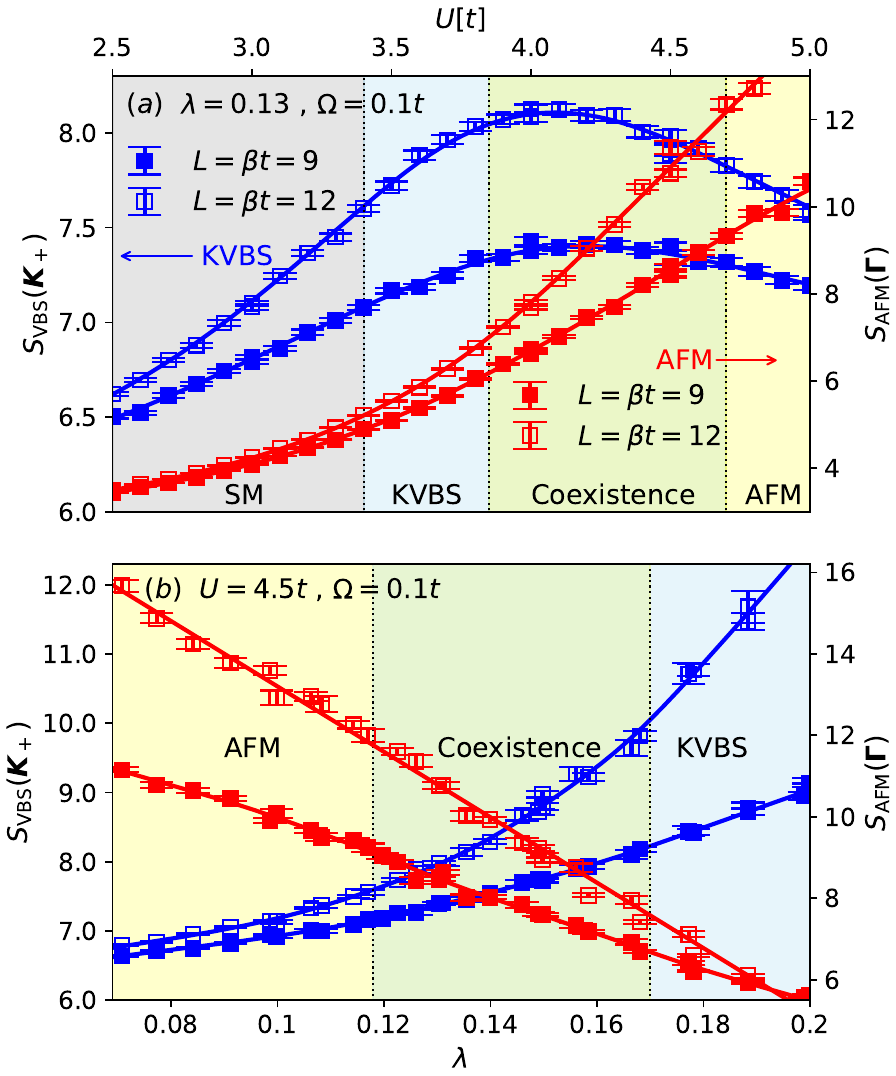}
    \vspace{-0.6cm}
    \caption{The evolution of the \gls*{KVBS}  $S_\mathrm{VBS}(\boldsymbol{K})$ and antiferromagnetic $S_\mathrm{AFM}(\boldsymbol{\Gamma})$ structure factors for $\Omega=0.1t$. Results are plotted as a function of (a) $U$ for constant $\lambda=0.13$ and (b) $\lambda$ for constant $U=4.5t$. The left [right] $y$-axis in both panels shows the $S_\mathrm{VBS}(\boldsymbol{ K})$ [$S_\mathrm{AFM}(\boldsymbol{\Gamma})$] values, as indicated by the arrows in the top panel. 
    The colored regions indicate the four distinct regions in the phase diagram, including the coexistence region, consistent with Fig.~\ref{fig:Phase_diagram}. The blue and red lines are guides for the eye.}
    \label{fig:Structurefactor_vs_couplings}
\end{figure}

\noindent\textbf{Results}.  Figure~\ref{fig:Phase_diagram} shows the ground state phase diagram for the half-filled \gls*{oSSH}-Hubbard model for $\Omega=0.1t$, and constitutes our main result. It hosts Dirac \gls*{SM}, \gls*{KVBS} and \gls*{AFM} phases. We also find evidence for a small region of \gls*{KVBS}-\gls*{AFM} coexistence, which may reflect a weakly first-order transition that is difficult to fully resolve in the simulations.\footnote{A similar coexistence region was also inferred for the square lattice Hubbard-\gls*{oSSH} model at finite temperatures~\cite{Tanjaroon2025Antiferromagnetic}.}
The \gls*{KVBS} phase appears for large $\lambda$ values, consistent with previous \gls*{QMC} investigations of the \gls*{oSSH}~\cite{Malkaruge2024Kekule} and \gls*{oSSH}-Hubbard model in which the \gls*{eph} interaction is treated at a mean-field level~\cite{Otsuka2024Kekule}. The \gls*{AFM} phase appears for $U \gtrsim 3.8t$. The critical coupling $U_\mathrm{c}$ defining the \gls*{AFM} phase boundary varies weakly with the \gls*{eph} coupling for small $\lambda$; however, once $U_c$ crosses the \gls*{KVBS} transition line it is rapidly pushed to larger values with increasing $\lambda$. Interestingly, the onset of the \gls*{AFM} phase is accompanied by a dip in \gls*{KVBS} transition line and the emergence of a coexistence region for $U/t \in (3.8, 6.0)$ and $\lambda \in (0.11, 0.18)$. Notably, the minimal $\lambda_\mathrm{c} \approx 0.11$ value occurs for a $U$ value that roughly coincides with the \gls*{QCP} of the honeycomb Hubbard model ($U_\mathrm{c} \approx 3.8t$~\cite{Sorella2012Absence, Assaad2013Pinning, Toldin2015Fermionic, Kennedy2025Extended}). We note that a similar dip in the \gls*{KVBS} $\lambda_c$ phase boundary in proximity to the onset of the \gls*{AFM} phase was reported in Ref.~\cite{Otsuka2024Kekule} when treating the lattice at the mean-field level. 

The red star in Fig~\ref{fig:Phase_diagram} is the approximate location of unstrained graphene in the phase diagram if we assume the relevant phonon modes are the in-plane bond stretching $A_1$ and $B_1$ modes~\cite{Malkaruge2024Kekule, Schuler2013Optimal, thePhysicsOfgraphene}; for more information refer to App.~\ref{sec:graphene}. While graphene  is placed in the \gls*{SM} region of the phase diagram, it is near the \gls*{SM}-\gls*{KVBS} phase boundary such that small perturbations that increase $\lambda$ could easily stabilize the \gls*{KVBS} phase. This proximity was also observed in our previous work investigating the $U=0$ limit of the \gls*{oSSH}-Hubbard model. Not only do our current results show that introducing a Hubbard interaction does not change this conclusion, but further suggest that increasing the effective $U$ may also stabilize the \gls*{KVBS} phase. This result has important implications for engineering and manipulating  graphene-based  systems.
\begin{figure}[t]
    \centering
    \includegraphics[width=0.80\linewidth]{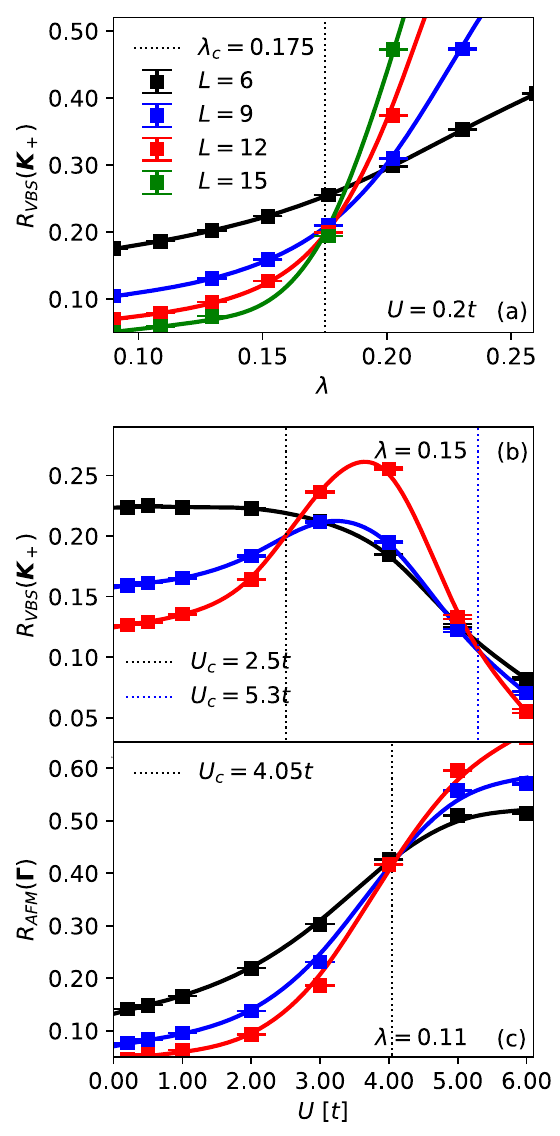}
    \caption{A crossing analysis using the correlation ratios $R_\text{VBS}(\boldsymbol{\boldsymbol{K}})$ and $R_\text{AFM}(\boldsymbol{\boldsymbol{\Gamma}})$ to determine the \gls*{QCP} for the \gls*{KVBS} and \gls*{AFM} phases at $\Omega=0.1t$. Here as we vary the lattice size, we set $\beta t =L$ to determine the location of the ground-sate phase boundaries. (a)  $R_\mathrm{VBS}(\boldsymbol{K})$ vs. $\lambda$ for fixed $U=0.2t$. The QCP for the SM-KVBS phase transition is observed at $\lambda_\mathrm{c} \approx 0.175$. (b) $R_\mathrm{VBS}(\boldsymbol{K})$ vs $U$ at $\lambda=0.15$. Two crossing points are observed at $U_\mathrm{c}\approx2.5t$ and $5.3t$, indicating two \gls*{QCP}s for this value of the $e$-ph coupling strength. (c) $R_\mathrm{AFM}(\boldsymbol{\Gamma})$ vs $U$ for a constant $\lambda=0.11$, with a QCP observed at $U \approx 4.05t$. Dashed lines in all panels indicate the estimated location of QCPs while the solid lines serve are guides to the eye.}
    \label{fig:corr_ratio_vs_couplings}
\end{figure}

We now discuss the numerical results used to determine the phase diagram. Fig.~\ref{fig:Structurefactor_vs_couplings} shows the structure factors $S_{\theta}(\boldsymbol{Q})$ as a function of $U$ for fixed $\lambda=0.13$ [panel (a)] and $\lambda$ for fixed $U=4.5t$ [panel (b)]. In both cases, we show results for $L=\beta t = 9$ and $12$. For $\lambda = 0.13$ (Fig.~\ref{fig:Structurefactor_vs_couplings}a), the system passes through the \gls*{SM}, \gls*{KVBS}, coexistence, and \gls*{AFM} regions of the phase diagram as $U$ increases.
$S_\text{AFM}(\boldsymbol{\Gamma})$ increases monotonically with $U$ along this cut through the phase diagram, while the $S_\mathrm{VBS}(\boldsymbol{K}_+)$ structure factor exhibits non-monotonic behavior and peaks in the coexistence region. Additionally, the $S_\mathrm{VBS}(\boldsymbol{K}_+)$ maximum shifts toward the boundary between the \gls*{KVBS} and coexistence region but remains within it with increasing lattice size $L$. This behavior suggests 
that the coexistence region may shrink in the thermodynamic limit, and could reflect the presence of a difficult to resolve weakly first-order phase transition. For larger $U$ values, this coexistence region disappears entirely and the \gls*{AFM} and \gls*{KVBS} transition lines merge on the clusters accessed here, consistent with a direct transition between the two phases in the strong coupling regime. 

Fig.~\ref{fig:Structurefactor_vs_couplings}b shows results for fixed $U=4.5t$ as a function of $\lambda$, where system passes through the \gls*{AFM}, coexistence, and \gls*{KVBS} phases, respectively, while entirely avoiding the \gls*{SM} region of the phase diagram. Along this cut $S_\mathrm{KVBS}(\boldsymbol{K}_+)$ monotonically increases and $S_\mathrm{AFM}(\boldsymbol{\Gamma})$ monotonically decreases with $\lambda$, indicating that increases to $\lambda$ effectively enhance \gls*{KVBS} correlations while suppressing \gls*{AFM} correlations. Additionally, both structure factors show a pronounced dependence on system size, not only within their respective ordered regions of the phase diagram but also in the coexistence region, consistent with a long-range ordered state. 

\begin{figure}[t]
    \centering
    \includegraphics[width=0.9\linewidth]{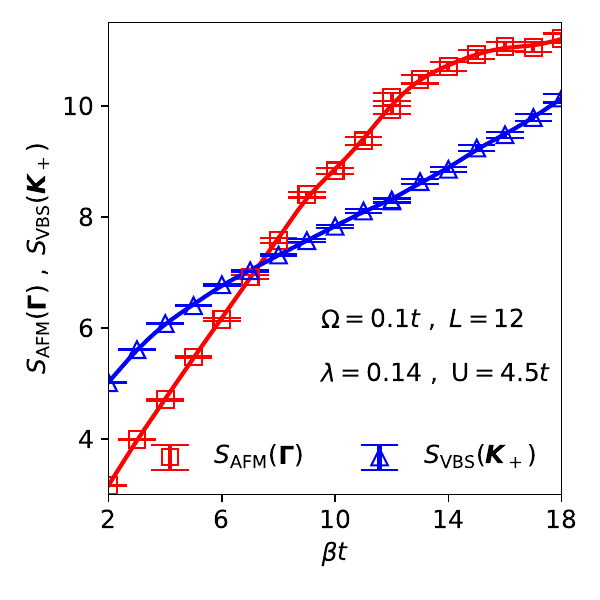}
    \vspace{-0.6cm}
    \caption{\gls*{AFM} $S_\mathrm{AFM}(\boldsymbol{\Gamma})$ (red $\Box$) 
    and \gls*{KVBS} $S_\mathrm{VBS}(\boldsymbol{K})$ (blue $\triangledown$) structure factors as a function of inverse temperature $\beta t$ for an $L=12$ lattice within the coexistence region of the phase diagram ($\Omega=0.1t$, $U=4.5t$, and $\lambda=0.14$). }
    \label{fig:T-dep}
\end{figure}

Figure~\ref{fig:corr_ratio_vs_couplings} provides representative examples of both the \gls*{AFM} and \gls*{KVBS} correlation ratio crossings used to determine the phase boundaries in Fig.~\ref{fig:Phase_diagram}. In all cases, we consider $L=\beta t = 6$, $9$, and $12$ but include $L=\beta t = 15$ when needed.  Fig~\ref{fig:corr_ratio_vs_couplings}a shows $R_\mathrm{VBS}(\boldsymbol{K}_+)$ as a function of $\lambda$ for $U=0.2t$, in which we obtain the three crossing points $\lambda_{c}^{L}$ for lattice size pairs ($L,L-3$). The crossing points are well converged for $\lambda_\mathrm{c}^{15}$; therefore, we take $\lambda_\mathrm{c}^{\text{KVBS}}\approx \lambda_\mathrm{c}^{15}=0.175$ as the critical coupling for the \gls*{KVBS} phase transition and estimate the corresponding uncertainty as $\Delta \lambda_\mathrm{c}^\text{KVBS} = \pm |\lambda_\mathrm{c}^{15}-\lambda_\mathrm{c}^{12}|=\pm 0.005$. We use this approach to obtain all \gls*{QCP}s. Fig.~\ref{fig:corr_ratio_vs_couplings}b similarly shows the results for increasing $U$ while keeping $\lambda=0.15$ fixed. Here we observe two crossing points, the first one corresponding to the \gls*{SM}/\gls*{KVBS} transition and the second corresponding to the \gls*{KVBS}/\gls*{AFM} transition. Lastly, Fig.~\ref{fig:corr_ratio_vs_couplings}c shows the crossing point $U^\text{AFM}_\mathrm{c}/t = 4.05 \pm 0.07$ for the \gls*{QCP} associated with an \gls*{AFM} transition at $\lambda = 0.15$.

Finally, Fig.~\ref{fig:T-dep} plots both the \gls*{KVBS} and \gls*{AFM} structure factors as a function of $\beta t$ within the coexistence region of the phase diagram ($L = 12$, $\lambda=0.14$ and $U=4.5t$). We observe that both structure factors increase monotonically as the temperature is lowered, an indication of phase coexistence. Furthermore, $S_\text{AFM}(\boldsymbol{\Gamma})$ grows more quickly at high temperatures before gradually saturating at lower temperatures. Conversely, $S_\text{VBS}(\boldsymbol{K}_+)$ initially grows more slowly at high temperatures, and then more rapidly at lower temperature. There is an inflection point at approximately $\beta t \sim 10$, where the curves switches from concave down to concave up with decreasing temperature. Whether this coexistence would persist in the thermodynamic limit or is simply a signature of weakly first order transition is difficult to confidently discern on the finite sized lattices we are able to simulate.\\

\noindent\textbf{Discussion}.  
Prior \gls*{DQMC} work on the \gls*{oSSH} honeycomb model established the presence of a \gls*{SM}/\gls*{KVBS} transition at $\lambda_\mathrm{c} \approx 0.18$ for quantum phonons and an energy $\Omega = 0.1t$~\cite{Malkaruge2024Kekule}. \gls*{QMC} studies have also identified a \gls*{SM}/\gls*{AFM} transition in the honeycomb Hubbard model 
at $U_\mathrm{c}\approx 3.8t$~\cite{Assaad2013Pinning, Otsuka2016Universal, Costa2021Magnetism}. Our  phase diagram in Fig.~\ref{fig:Phase_diagram} is quantitatively consistent with these results, while also qualitatively resembling the phase diagram predicted in Ref.~\cite{Otsuka2024Kekule}, which treated the phonons at the mean-field level. Interestingly, while $U_\mathrm{c}$ increases monotonically with $\lambda$, $\lambda_\mathrm{c}$ has a nonmonotonic dependence on $U$, with $\lambda_\mathrm{c}$ minimized for a Hubbard repulsions close to $U \approx 3.8t$. More specifically, introducing a small Hubbard interaction  $U \lesssim 3.8$ effectively enhances the \gls*{KVBS} correlations. This is likely a result of the Hubbard interaction suppressing double occupancy and promoting short-range \gls*{AFM} correlations that favor local singlet formation consistent with the \gls*{KVBS} phase. We also observe indications of an \gls*{AFM}-\gls*{KVBS} coexistence region within a relatively small window $3.8t \le U < 6t$. Whether this coexistence region persists in the thermodynamic limit, or is a result of finite size effects associated with a weakly first order transition, is difficult to discern at the lattice sizes we can access in our simulations.

First principles calculations place graphene in the \gls*{SM} region of the phase diagram but near \gls*{SM}-\gls*{KVBS} phase boundary. This location suggests that the system will be very sensitive to any external perturbations that increase either the effective on-site Hubbard repulsion or the strength of the \gls*{eph} interaction. Moreover, changes in these parameter are likely to drive graphene-based systems into the \gls*{KVBS} as opposed to \gls*{AFM} insulating phase. This result is consistent with experimental observations of strained graphene systems, where Kekul{\'e}-O lattice distortions are frequently observed~\cite{Zhang2022Self}. Our results suggest that this phase could be manipulated either through strain engineering, modifying electronic screening channels, or pump-probe approaches to modifying $U$~\cite{TancogneDejean2018ultrafast, Baykusheva2022ultrafast, Granas2022ultrafast}.  \\

\noindent\textbf{Acknowledgments} --- This work was supported by the National Science Foundation under Grant No. DMR-2401388.

\appendix
\section{Simulation parameters}\label{sec:dqmc_details}
We perform simulations on clusters ranging in size from $L=6 \text{ to }15$. The number of orbitals for a given lattice size is $N = 2L^2$. We typically performed an initial $\sim 5\times 10^3$ \gls*{HMC} updates to thermalize the system, followed by an additional $\sim 1\times 10^4$ updates, after each of which measurements were made. All simulations ran $8-12$ parallel Markov chains to improve the measurement statistics. Throughout all the \gls*{DQMC} simulations, the imaginary-time discretization constant was set to $\Delta\tau=0.05$. For the \gls*{HMC} updates, $N_t \sim 10$ time steps were performed, with the corresponding time-step size given by $\Delta t = \pi/(2\Omega N_t)$. 

Throughout this work, we monitored the number of times that the \gls*{SSH} coupling results in an inversion of the sign of the hopping and ensured that it remained in the physical regime where the linear approximation is valid (i.e., below $1\%$)~\cite{Banerjee2023groundstate, Malkaruge2024Kekule}.\\

\section{Parameter Estimates for Graphene}\label{sec:graphene}
The phonon modes in graphene that are thought to play an important role in the formation of \gls*{KVBS} correlations are the in-plane bond-stretching $A_1$ and $B_1$ optical modes, with energy $\Omega \approx 170\text{meV} =  0.065t$~\cite{Wu2018Theory,Basko2007Effect,Basko2008Interplay}. In Ref.~\cite{Malkaruge2024Kekule} we used \textit{ab initio} calculations from Ref.~\cite{Ribeiro2009Strained} to estimate that microscopic \gls*{eph} coupling is $\alpha \approx 6.09~\text{eV/\AA}$, with corresponding dimensionless \gls*{eph} coupling $\lambda \approx 0.17$. The phonon energy $\Omega = 0.1t$ used in this work is in the adiabatic regime relevant to graphene, and relatively close in value to the $A_1$ and $B_1$ energy. Additionally, in the adiabatic limit, the \gls*{KVBS} phase boundary is primarily controlled by $\lambda$, and is largely insensitive to the precise value of the phonon energy, provided that $\Omega/t \ll 1$~\cite{Malkaruge2024Kekule}. The effective Hubbard interaction $U \approx 1.6t$ for graphene is taken from Ref.~\cite{Schuler2013Optimal}, which used the Peierls-Feynman-Bogoliubov variational principle, in conjunction with comparisons to \gls*{DQMC} simulation results, to estimate a renormalized local Hubbard interaction that accounts for the effects of nonlocal Coloumb interactions.

\bibliography{references.bib}
\end{document}